\documentclass[prd,12pt,tightenlines,onecolumn,amssymb,preprintnumbers,nofootinbib,superscriptaddress,]{revtex4-2}

\pdfoutput=1

\usepackage{hyperref}
\usepackage{feynmf}
\usepackage{graphicx}
\usepackage{enumitem}
\usepackage{latexsym}
\usepackage{amsfonts}
\usepackage{amssymb}
\usepackage{mathrsfs}
\usepackage{color}
\usepackage{amsmath}
\usepackage{slashed}
\usepackage{dcolumn}
\usepackage{verbatim}
\usepackage{comment}
\usepackage{float}
\usepackage{multirow}
\usepackage{xspace}
\usepackage[normalem]{ulem}

\newcommand{\eV}{~\text{eV}}

\newcommand{\MeV}{~\text{MeV}}
\newcommand{\GeV}{~\text{GeV}}

\newcommand{\Neff}{\Delta N_\text{eff}}

\begin{document}

\title{Signatures of Bulk Neutrinos in the Early Universe}

\author{David McKeen}
\email{mckeen@triumf.ca}
\affiliation{TRIUMF, 4004 Wesbrook Mall, Vancouver, BC V6T 2A3, Canada}

\author{John N. Ng}
\email{misery@triumf.ca}
\affiliation{TRIUMF, 4004 Wesbrook Mall, Vancouver, BC V6T 2A3, Canada}

\author{Michael Shamma}
\email{mshamma@triumf.ca}
\affiliation{TRIUMF, 4004 Wesbrook Mall, Vancouver, BC V6T 2A3, Canada}

\begin{abstract}
Neutrino masses and quantum gravity are strong reasons to extend the standard model of particle physics. A large extra dimension can be motivated by quantum gravity and can explain the small neutrino masses with new singlet states that propagate in the bulk. In such a case, a Kaluza-Klein tower of sterile neutrinos emerges. We revisit constraints on towers of sterile neutrinos that come from cosmological observables such as the effective number of noninteracting relativistic species and the dark matter density. These limits generically rule out micron-sized extra dimensions. We explore the weakening of these constraints to accommodate an extra dimension close to the micron size by assuming that the universe reheated after inflation to a low temperature. We discuss how such a possibility can be distinguished in the event of a positive signal in a cosmological observable.
\end{abstract}
\maketitle

\section{Introduction}

The observation of neutrino flavor oscillations in astrophysical and terrestrial neutrino experiments is unambiguous evidence of nonzero neutrino masses~\cite{Pontecorvo:1957cp,Pontecorvo:1967fh,KamLAND:2008dgz}.
In order to accommodate this observation, the standard model (SM) must be extended since one cannot construct a gauge-invariant dimension-4 mass term for the neutrino with SM fields alone.
The most minimal model that gives all three neutrino masses extends the SM with three color and hypercharge singlet, right-handed (or ``sterile'') fields $\nu_R$.
Each singlet couples to the SM through the operator $y\bar{L}H\nu_R$ so that after electroweak symmetry breaking, each neutrino obtains a Dirac mass $m_\nu\sim yv$ where $v$ is the Higgs vacuum expectation value.
To obtain eV-scale neutrino masses requires $y\sim10^{-12}-10^{-11}$ which makes terrestrial and cosmological production of the sterile states exceedingly improbable. 
However, new interactions beyond this could lead to additional channels of production for the sterile neutrino and rich phenomenological consequences at neutrino experiments on (under) the ground and in the sky~\cite{Abazajian:2019oqj,Adshead:2020ekg,Luo:2020sho,Luo:2020fdt,Mahanta:2021plx,Du:2021idh,Heeck:2023soj,King:2023cgv,Cox:2023dou,Batell:2016zod}.

At the same time there are theoretical reasons to think that the SM should be extended. First, the SM lacks a quantum mechanical description of gravitational interactions. 
Formulating a consistent quantum description of gravity has been the main objective of string theory~\cite{Ibanez:2012zz,Cvetic:2022fnv}.
Much effort has been expended on appropriate low-energy compactifications of the extra dimensions in string theory that produce the four dimensional SM with gravity.
In particular, attention has been recently focused on low energy effective field theories that are and are not consistent with quantum gravity, referred to as the ``string theory landscape'' and ``swampland'', respectively (see e.g.~\cite{Brennan:2017rbf,Palti:2019pca,Agmon:2022thq,Vafa:2024fpx} for recent reviews).
Motivated by these conjectures in string theory, a micron-sized dark dimension, or large extra dimension (LED), has been proposed to explain the small cosmological constant and includes a graviton whose tower of Kaluza-Klein (KK) modes may comprise (some of) the dark matter (DM)~\cite{Montero:2022prj,Schwarz:2024tet,Obied:2023clp,Law-Smith:2023czn,Gonzalo:2022jac}.
The dark dimension scenario can only accommodate SM singlet fields so additional contributions to the DM may include KK excitations of a bulk neutrino.
In fact, the dark dimension also presents an elegant explanation of the smallness of neutrino masses~\cite{Arkani-Hamed:1998wuz,Dienes:1998sb,Mohapatra:2000wn,Anchordoqui:2022svl}. 
In LED models of neutrino mass, new SM gauge singlet right-handed neutrinos, fields propagate in the $4+n$ dimensional space called the ``bulk'', where $n$ are the number of additional spatial dimensions. 
On the other hand, the fields which have SM charges are confined to propagate in the 4D ``brane''. 
Yukawa couplings of the SM (active) neutrinos $\nu_L$, SM Higgs $H$, and the bulk neutrinos give rise to Dirac neutrino masses through the familiar Higgs mechanism at the weak scale. 
The neutrino mass is suppressed not by tiny couplings in this case, but by the volume of the extra dimensions.

Large extra dimensions can be investigated in terrestrial experiments via their instigation of short-baseline neutrino oscillations. 
Additionally, the extra dimension can be explored via the perturbations they generate in solar, atmospheric, and long-baseline neutrino experiments.
Specifically, the LEDs are probed through the effects of the Kaluza-Klein (KK) modes which describe the right-handed singlet neutrino fields on the brane.
Given the micron size of the LED, the observable effect at oscillation experiments is quite similar to that of eV-scale sterile neutrinos.
When connecting to these experiments, it is typically sufficient to consider the effects of the largest LED~\cite{McLaughlin:2000zf,McLaughlin:2000iq,Machado:2011jt,Machado:2011kt,Basto-Gonzalez:2012nel,Girardi:2014gna,Stenico:2018jpl,Forero:2022skg}.
In this situation, the model is described by two parameters: the (active) neutrino mass scale $m$ and the radius of the extra dimension $R$.

In this work, we point out that the KK modes of a bulk neutrino are produced abundantly in the early universe and in turn may be observed or constrained by ongoing and upcoming cosmological surveys.
In particular, these modes contribute to the DM density and to the relativistic energy density at the time of recombination.
Whether a given KK mode contributes to the DM density or to relativistic energy is determined by its mass and its lifetime, factors which depend on its position in the KK tower and the size of the extra dimension.\footnote{Interestingly, this setup can be thought of as an example of ``dynamical dark matter''~\cite{Dienes:2011ja,Dienes:2011sa} in which a large range of dark states with varying lifetimes can have novel phenomenological consequences.}
Crucially, the number of bulk neutrino modes which are produced depends on the temperature at which the universe reheats with the production of heavier modes exponentially suppressed.
If the universe reheats to very high temperatures, then higher states in the KK tower can be produced and drastically affect cosmological observables~\cite{Abazajian:2000hw}.
On the other hand, there has been model building effort focused on the possibility of constructing the observed universe at $T\sim\mathcal{O}(10\MeV)$~\cite{Dimopoulos:1987rk}, especially recently~\cite{Ghalsasi:2015mxa,Aitken:2017wie,Elor:2018twp,Nelson:2019fln,Alonso-Alvarez:2019fym,Elor:2020tkc,Bhattiprolu:2022sdd,Berger:2023ccd,Silva-Malpartida:2023yks, Bernal:2023ura}, and low reheat temperatures can lessen constraints on sterile neutrinos~\cite{Abazajian:2023reo}.
In this work we not only recast cosmological constraints on models of extra dimensional neutrinos in the case of high reheating temperatures but also provide a fresh analysis of cosmological implications of this model if the universe was only as hot as a few$\MeV$.

The paper is organized as follows. In Sec. \ref{sec:exdim} we review the necessary features for neutrino mass generation in models of LEDs. 
Then, in Sec. \ref{sec:probes}, we review how sterile neutrinos are probed in terrestrial experiments.
Next, in Sec. \ref{sec:cosmoprobes} we review how beyond the Standard Model (BSM) physics affects $\Lambda$CDM cosmology and study in detail the effects of LED models of neutrino masses on cosmological observables.
We conclude this work in Sec. \ref{sec:conc} with a summary of our main results.

\section{Neutrinos in Large Extra Dimensions}\label{sec:exdim}

Our starting point is an extra dimensional model in which all SM fields are localized on a single 4D brane while SM singlet fields are free to propagate in the extra dimensions. In particular, we allow sterile neutrinos to be bulk fields. The ingredients in our model are SM lepton doublets $L_\alpha=(\nu_{\alpha L},\ell_{\alpha L})^T$ with $\alpha=e,\mu,\tau$ labeling the lepton flavour bulk fermions $N_\alpha$.\footnote{To give the three active neutrinos masses we have added three sterile neutrino fields.} We are interested in the situation where one of the extra dimensions, denoted by $y$, is compactified onto a circle of radius $R$ which is much larger than the sizes of the other extra dimensions~\cite{Machado:2011jt,Machado:2011kt,Basto-Gonzalez:2012nel,Girardi:2014gna,Stenico:2018jpl}. The Bulk fermions satisfy the boundary condition $N_\alpha(x,y)=N_\alpha(x,y+2\pi R)$. The relevant terms in the 5D action, assuming the interactions of the SM fields take place on the SM brane at $y=0$, are
\begin{equation}
\begin{aligned}
S&=\int d^4xdy \bar{N}_\alpha i\Gamma^AD_AN_\alpha+\int d^4x\bar{L}_{\alpha} i\gamma^\mu\partial_\mu L_{\alpha}\\ 
&\quad\quad\quad\quad\quad\quad-\frac{\lambda_{\alpha\beta}}{\sqrt{M_\ast}}\int d^4x dy\bar{L}_{\alpha}(x)N_\beta(x,y)H(x)\delta(y)+\text{h.c.}
\end{aligned}
\label{eq:5daction}
\end{equation}
where $\Gamma^A=(\gamma^\mu,i\gamma^5)$, $D_A$ is the 5D partial derivative operator, and $H$ is the SM Higgs fields. The Yukawa couplings $\lambda_{\alpha\beta}$ are dimensionless and we introduced $M_\ast$, which is the scale at which this extra dimensional description breaks down. Although there are a number of interpretations of $M_*$, we simply assume $M_*\gg M_\text{EW}$ where $M_\text{EW}$ is the electroweak energy scale.

We perform a Kaluza-Klein decomposition of the extra dimensional fields and set $H=(v/\sqrt2,0)^T$ where $v=246~\GeV$ is the Higgs vacuum expectation value. Doing so generates kinetic terms for the active neutrinos $\nu_{\alpha L}$ and infinite towers of bulk neutrinos $n_{\alpha L,R}^{\pm k}$ with $k=1,2,\dots$ labelling the state. In addition, the Yukawa interaction generates interaction terms between the bulk and active neutrinos,
\begin{equation}
\begin{aligned}
S&\supset-\int d^4x\bigg\{\sum_{k=1}^{\infty}\bigg(m_k\bar{n}_{\alpha L}^kn_{\alpha R}^k+m_{-k}\bar{n}_{\alpha L}^{-k}n_{\alpha R}^{-k}\bigg)
\\
&\quad\quad\quad\quad\quad\quad\quad\quad+m^D_{\alpha\beta}\bar{\nu}_{\alpha L}\bigg[n_{\beta R}^0+\sum_{k=1}^{\infty}\big(n_{\beta R}^k+n_{\beta R}^{-k}\big)\bigg]+\text{h.c.}\bigg\}
\end{aligned}
\label{eq:effaction}
\end{equation}
where $m_k=-m_{-k}=k/R$. The first term above represents masses for Dirac fermions made up of bulk neutrinos while the second involves Dirac masses that mix the active neutrinos $\nu_{\alpha L}$ with the bulk neutrinos,
\begin{equation}\label{eq:diracmass}
m_{\alpha\beta}^D=\frac{\lambda_{\alpha\beta} v}{\sqrt{4\pi M_*R}}.
\end{equation}

We can also relate the scale $M_*$ to the gravitational scale through the relation~\cite{Arkani-Hamed:1998wuz,Dienes:1998sb,Mohapatra:2000wn}
\begin{equation}\label{eq:newscale}
\bigg(\frac{M_\text{Pl}}{M_*}\bigg)^2=M_*(2\pi R).
\end{equation}
One can check for what values of $\lambda,~M_*$ reproduce neutrino masses roughly on the order of current bounds.
For example, with $\lambda=0.1$, $M_*\sim10^8\GeV$ the neutrino mass is $m^D\sim0.1\eV$. 
Hence small Dirac neutrino masses are natural in the LED scenario. In this minimal model the neutrinos are Dirac and lepton number is conserved; to realize Majorana masses for the neutrinos in extradimensional scenarios see, e.g.~\cite{Pilaftsis:1999jk}.

The interactions in Eq.~(\ref{eq:effaction}) give rise to a mass matrix which can be diagonalized by a unitary transformation. After diagonalization, the mixing of the three active neutrino flavor eigenstates can be decomposed into mass eigenstates, $\nu_i^{(j)}$, as
\begin{equation}\label{eq:numixing}
    \nu_{\alpha L}=\sum_{i=1}^3U_{\alpha i}\sum_{j=0}^\infty V_{ij}\nu_{i}^{(j)},
\end{equation}
where $U_{\alpha i}$ is the $3\times3$ Pontecorvo-Maki-Nakagawa-Sakata (PMNS) matrix describing the mixing of the active neutrino flavours. The three KK towers correspond to the three light neutrino masses observed in oscillation data. The $\nu_i^{(k)}$ have $k=0$ modes that are mostly active while $k>0$ modes correspond to bulk neutrino modes with small active admixture. The masses of the neutrinos, $m_i^{(k)}$, are determined by the eigenvalue equation
\begin{equation}
m_i^{(k)}R-( m_i^D R)^2\pi\cot(\pi m_i^{(j)}R)=0
\end{equation}
where $m_i^D$ are the eigenvalues of the Dirac neutrino mass matrix. 
In the above expression  We observe that the usual mass eigenstates are each accompanied by an infinite tower of bulk, or sterile, neutrinos. The admixture of these heavy neutrinos with the active neutrinos is controlled by the components of the matrix $V_{ij}$ are determined by~\cite{Dienes:1998sb,Mohapatra:2000wn}
\begin{equation}\label{eq:actstermix}
    V_{ij}^2=\frac{2}{1+\pi^2(m_i^DR)^2+(m_i^{(j)}/m_i^D)^2}
\end{equation}
The transcendental equation can be analytically solved when $m_i^DR\ll1$. In this limit,
\begin{align}
    m_i&\equiv m_i^{(0)}=m_i^D\bigg[1-\frac{\pi^2}{6}(m_i^DR)^2+\dots\bigg],\label{eq:k0mass}\\ M_{ik}&\equiv m_i^{(k)}=\frac{k}{R}\bigg[1+\left(\frac{m_i^DR}{k}\right)^2+\dots\bigg],~{\rm for}~k>0, \label{eq:kg0mass}\\ V_{i0}&=1-\frac{\pi^2}{6}(m_i^DR)^2+\dots, \label{eq:k0mix}\\ V_{ik}&=\frac{\sqrt{2}m_i^DR}{k}\bigg[1-\frac{3}{2}\frac{(m_i^DR)^2}{k^2}+\dots\bigg],~{\rm for}~k>0 .\label{eq:kg0mix}
\end{align}
For the $k>0$ modes, the masses increase and the mixings with active flavours $V_{ik}$ decrease for increasing $k$. We note also that the mixings between the sterile neutrinos are parametrically smaller than the active-sterile mixings by an additional factor of $m_i^DR$. 
In the parameter space we are interested in, $m_i^DR\ll1$ so that $m_i\simeq m_i^D$ and $M_{ik}\simeq k/R$. Thus, each of the three KK modes with the same value of $k>0$ are nearly degenerate. In what follows, we will refer to their common mass without the zero mode label, $M_{ik}\equiv M_k$ for $i=1,2,3$.

Consequences of the interactions between the sterile and active neutrinos given in Eq. \ref{eq:numixing} include the ability for SM neutrinos to oscillate into $k\neq0$ sterile neutrinos and, as we will see, production $k>0$ modes in the early universe. 

\section{Probes of Sterile Neutrinos}\label{sec:probes}

The main focus of this work will be on the effects of the production of KK-modes in the early universe.
However, it will be important to compare the reach of cosmological experiments to terrestrial experiments in their ability to observe or constrain an extra dimensional scenario.
In this section we review some of the terrestrial experiments which can observe and constrain sterile neutrinos with a particular focus on how the excitations of bulk neutrinos can be observed.

\subsection{Terrestrial Probes of Sterile Neutrinos}\label{sec:terrestrial}
It behooves us to begin our discussion with the connection to searches for sterile neutrinos at oscillation experiments. As is well known, because neutrinos have mass, their mass eigenstates do not correspond to the states that couple to the charged leptons, i.e. lepton flavours, through weak interactions. This leads to neutrino flavour oscillations while they propagate between production and detection via charged current weak interaction.

In the dark dimension scenario, the charged leptons couple dominantly to the zero modes of the KK neutrino towers but have nonzero couplings to each of the $k>0$ modes. The probability to produce a neutrino of flavour $\alpha$ and with energy $E$ and measure it as flavour $\beta$ after traveling a distance $L$ can be written as
\begin{equation}\label{eq:oscprob}
P(\nu_\alpha\rightarrow\nu_\beta)=\bigg|\sum_{i=1}^3\sum_{k=0}^{\infty}U_{\alpha i}^*U_{\beta i}V_{ik}^2\exp\bigg(-i\frac{(m_i^{k})^2L}{2E}\bigg)\bigg|^2.
\end{equation}
Since we are working in the part of parameter space with $R\ll 1/m_i$, oscillations amongst active neutrinos are not grossly changed from the SM scenario; solar, atmospheric, and reactor neutrino oscillation data can be fit by splittings between the mostly active zero-mode neutrinos determined by $\Delta m_{21}^2=\Delta m^2_\odot\simeq 7.5\times10^{-5}~{\rm eV}^2$, $\left|\Delta m_{31}^2\right|=\Delta m^2_{\rm atm}\simeq2.5\times10^{-3}~{\rm eV}^2$, and $U$ the usual (nearly) unitary $3\times 3$ PMNS matrix~\cite{Pontecorvo:1957qd,Maki:1962mu} ,
\begin{align}
    U=\begin{pmatrix}U_{e1} & U_{e2} & U_{e3}\\ U_{\mu 1} & U_{\mu 2} & U_{\mu 3}\\ U_{\tau 1} & U_{\tau 2} & U_{\tau 3}\end{pmatrix} \approx\begin{pmatrix}0.81 & 0.56 & 0.15\\ -0.47 & 0.49 & 0.73\\ 0.34 & -0.66 & 0.67\end{pmatrix}.
\end{align}
The numerical values here are the central values reported in~\cite{Esteban:2020cvm} for a normal ordered mass hierarchy and ignoring the so far unmeasured Dirac Charge-Parity (\textit{CP}) phase. An inverted hierarchy gives very slightly different values.

The presence of more than three neutrinos in the expression in Eq.~(\ref{eq:oscprob}) can alter the pattern of neutrino oscillations from the usual SM picture. Because the coupling to mode $k$ decreases with increasing $k$, we can focus on the mixing with the $k=1$ modes. Moreover, the near degeneracy among the $k=1$ states simplifies the analysis. In particular, for $R\sim{\cal O}(\mu{\rm m})$, experiments sensitive to $\rm eV$-scale sterile neutrinos can impact this scenario.

In the $m_i^DR\ll 1$ limit, active-sterile mass splittings are $\Delta m^2=M_1^2-m_i^2\simeq M_1^2=1/R^2$ for $i=1,2,3$ and the probability that an electron neutrino remains an electron neutrino can be written
\begin{align}\label{eq:appear}
P_{e\rightarrow e}=1-\sin^22\theta_{ee}\sin^2\bigg(\frac{M_1^2L}{4E}\bigg),
\end{align}
with
\begin{equation}
    \sin^22\theta_{ee}\simeq\frac{8}{M_1^2}\sum_{i=1}^3m_i^2|U_{ei}|^2.
\end{equation}\label{eq:eemixing}
In the limit that the light, mostly active neutrinos are degenerate, the approximate unitarity of the PMNS matrix simplifies this mixing angle considerably, $\sin^22\theta_{ee}\simeq 8 m^2R^2$ with
\begin{equation}
\begin{aligned}
m\equiv\left(\sum_{i=1}^3m_i^2\right)^{1/2}.
\end{aligned}
\label{eq:msq}
\end{equation}

\begin{figure}[t]
    \includegraphics[width=0.7\linewidth]{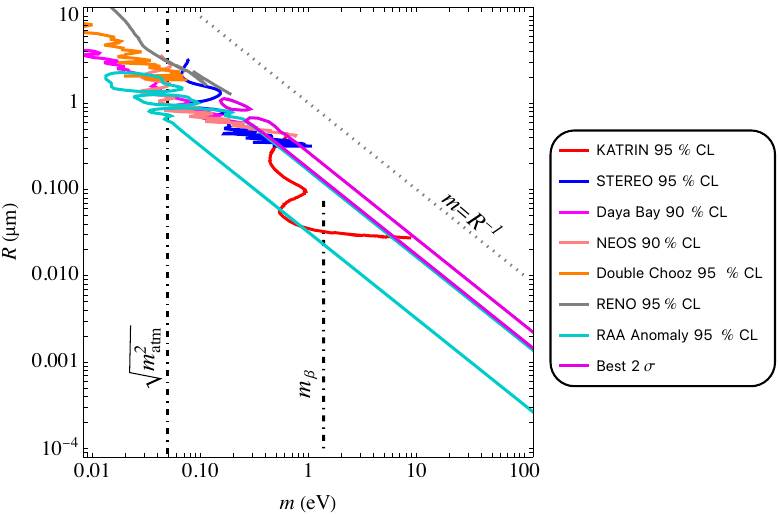}~
    \caption{Constraints on $m$, defined in Eq.~(\ref{eq:msq}), and $R$ from terrestrial experiments KATRIN~\cite{KATRIN:2022ith}, STEREO~\cite{STEREO:2019ztb}, Daya Bay~\cite{MINOS:2020iqj}, NEOS~\cite{NEOS:2016wee}, Double Chooz~\cite{DoubleChooz:2020pnv}, and RENO~\cite{RENO:2020hva}. We also show the favored regions for explaining the reactor anti-neutrino anomaly (RAA)~\cite{Mention:2011rk} and BEST~\cite{Barinov:2022wfh} results. The lower limit on $m$ from atmospheric neutrino oscillations~\cite{Super-Kamiokande:1998kpq} and as well as the upper bound on $m$ from the limit on the effective electron neutrino mass in tritium beta decay~\cite{KATRIN:2021uub} are also displayed (black, dot-dashed). Also shown is the boundary of the large active-sterile mass splitting regime, $m=R^{-1}$, above which higher order terms in Eqs.~(\ref{eq:k0mass})-(\ref{eq:kg0mix}) cannot be neglected.}
    \label{fig:terrestrial}
\end{figure}

In Fig.~\ref{fig:terrestrial}, we show constraints and favoured regions in the parameter space of $R$ and $m$ from experiments sensitive to electron neutrino disappearance due to ${\cal O}({\rm eV})$ sterile neutrinos. We assume that the light neutrinos are degenerate $m_1=m_2=m_3$ and consider only active-sterile oscillations into the first KK mode. The lower bound $m>\sqrt{\Delta m^2_{\rm atm}}$ required to accommodate atmospheric neutrino oscillations~\cite{Super-Kamiokande:1998kpq,MACRO:2001fie,Soudan2:2003qqa} and the upper bound $m/\sqrt3< m_\beta=0.8~\rm eV$ from measurements of the tritium $\beta$-decay endpoint~\cite{KATRIN:2021uub} are plotted. Active neutrino oscillations impose the lower bound $m^2\gtrsim\Delta m^2_{\rm atm}$ which we also indicate. While some hints for sterile neutrinos in this mass range can be accommodated with $R\sim(0.1-1~\mu{\rm m})$, we will see in Sec.~\ref{sec:cosmoprobes} that cosmological constraints generally preclude this region unless the standard cosmological history is changed quite drastically.

Additionally, we mention in passing that active-active oscillations such as $\nu_\mu\to\nu_e$ are parametrically suppressed in this setup as is typical of $3+1$ setups which this essentially is~\cite{Bilenky:1996rw,Okada:1996kw,Kopp:2013vaa}.

Lastly, there are bounds from charged lepton flavor violating processes that, because of the ``nondecoupling'' of the KK tower of sterile neutrinos can be rather strong~\cite{Ioannisian:1999cw}. However, the precise location of these bounds depends on the UV details of the extradimensional theory, such as the value of $M_*$, and we conservatively consider limits that are UV-insensitive.

\section{Cosmological Probes of Sterile Neutrinos}\label{sec:cosmoprobes}

Depending on their properties such as whether they are relativistic or non-relativistic and if they come into chemical equilibrium with the SM bath, the spectrum of KK-modes contribute to the radiation energy density or DM energy density.
Before we conduct a detailed study of these effects, we review how the contributions of additional degrees of freedom lead to changes in $\Lambda$-cold dark matter ($\Lambda$CDM) cosmology.

In this section we will describe the constraints on extradimensional models of neutrino masses arising in cosmology. These constraints can be quite strong because a large amount of the neutrino KK towers can be appreciably produced even if their mixing angles are small, i.e. even for small $R$ and large $k$ in Eq.~(\ref{eq:kg0mix}). 

In~\ref{sec:sterileprod}, we estimate the number density of each sterile mode produced through ``freeze in'' via small couplings to weak interactions. For the parameter space we are interested in, most of these states are nonrelativistic after the epoch of primordial nucleosynthesis and can, depending on their lifetimes, contribute to the dark matter density or act as a decaying dark matter component as we estimate in~\ref{sec:dm}. We will see in~\ref{sec:Neff} that, although they behave as matter, very strong constraints come from the contribution to the effective number of relativistic degrees of freedom when they decay and increase the number of light SM neutrinos. This is particularly important given near-future cosmic microwave background (CMB) observations that will increase these constraints or discover the effects of the heavy KK towers of sterile neutrinos.

\subsection{Sterile neutrino production in the early universe}\label{sec:sterileprod}
Despite their small couplings to SM states, the sterile neutrinos can be produced in appreciable quantities in the early universe. 
The production of the heavier, mostly sterile neutrinos comes through weak interactions controlled by their small active admixture.\footnote{Although we do not explore this possibility, allowing the sterile neutrinos to couple to other bosons can qualitatively change this picture; see, e.g.,~\cite{DeGouvea:2019wpf,Kelly:2020pcy}.}

In the part of parameter space that we are interested in, $m_i^D R\ll1$. We can therefore identify each of the three light neutrino masses with the Dirac mass parameters in the Lagrangian, $m_i=m_i^D$, and ignore the mass splittings between the three mostly sterile neutrinos corresponding to each $k>0$. Thus, for a fixed $k=1,2,\dots$, each of the three states $\nu_i^{(k)}$ can be identified with a single state labelled with mass $M_k=k/R$. If we make the assumption that the differences among the neutrino flavors in the early Universe is negligible (which is good to the ${\cal O}(1)$ level, sufficient for our purposes), then the production of each of these $k$ states in the early Universe is controlled by the square of the active-sterile mixing $\theta_k\simeq\sqrt{2}mR/k$ with $m$ defined in Eq.~(\ref{eq:msq}). The number density of $\nu_k$ evolves with time according to
\begin{equation}
\begin{aligned}
\frac{dn_k}{dt}+3Hn_k&=\frac14 \sin^22\theta_{k,\rm eff}\Gamma_\nu n_\nu
\end{aligned}
\label{eq:dndt}
\end{equation}
In this expression, $H$ is the Hubble expansion rate, $n_\nu$ is the number density of an active neutrino species, $\Gamma_\nu$ is the production rate of that neutrino, and $\theta_{k,{\rm eff}}$ is effective active-sterile mixing element in the presence of a nontrivial density of SM particles. Generically, cosmological limits preclude the sterile neutrinos from achieving chemical equilibrium with the rest of the SM plasma, so we focus on their production through ``freeze in,'' ignoring a depletion term on the right-hand side of Eq.~(\ref{eq:dndt}).

The effective active-sterile mixing angle can be described in terms of the density of the SM plasma, or equivalently, its temperature $T$,\footnote{We use $T$ to refer to the temperature of the active neutrinos. Before neutrino decoupling at $T\simeq3~{\rm MeV}$ this is the same temperature as the rest of the SM plasma. Afterwards, $e^+e^-$ annihilations reheat the photons to $T_\gamma=(11/4)^{1/3}T$.}
\begin{equation}
\begin{aligned}
\sin2\theta_{k,\rm eff}=\frac{\sin2\theta_k}{1+(T/T_V^k)^6}
\end{aligned}
\end{equation}
where~\cite{Dodelson:1993je,Abazajian:2000hw,Abazajian:2005gj}
\begin{equation}
\begin{aligned}
T_V^k&\sim250~{\rm MeV}\left(\frac{M_k}{1~\rm keV}\right)^{1/3}\simeq 320~{\rm MeV}\times k^{1/3}\left(\frac{10^{-4}\mu\rm m}{R}\right)^{1/3}.
\end{aligned}
\end{equation}
At temperatures below the weak scale, $T<T_w\sim100~\rm GeV$, $\Gamma_\nu= A G_F^2T^5$. We ignore the flavor-dependence of the ${\cal O}(1)$ prefactor $A$ in this rate and fix it to $A=1$ since the ${\cal O}(1)$ error that doing so introduces is subleading. For $T>T_w$, $\Gamma_\nu\propto T$ although we will see that production above this temperature is negligible---equivalently production of modes with mass $M_k\gtrsim T_w$ is very subleading.

Rewriting Eq.~(\ref{eq:dndt}) in terms of yields, $Y_{k,\nu}=n_{k,\nu}/s$, where $s$ is the entropy density of the Universe, to express the sterile neutrino production as a function of temperature gives
\begin{equation}
\begin{aligned}
\frac{dY_k}{dT}&=-\frac14 \frac{\sin^22\theta_k}{\left[1+(T/T_V^k)^6\right]^2}\frac{\Gamma_\nu Y_\nu}{HT}.
\end{aligned}
\label{eq:dYdT}
\end{equation}
In all of the parameter space that we are interested in, the production of the sterile states occurs in a radiation-dominated Universe, with $H=1.66\sqrt{g_\star(T)}T^2/M_{\rm Pl}$ with $g_\star(T)$ the effective number of relativistic degrees of freedom, and $M_{\rm Pl}=1.2\times10^{19}~\rm GeV$ the Planck mass. In what follows, we will investigate two scenarios: (i) a high-reheat situation where the SM plasma is reheated after inflation to a temperature well above the weak scale, $T_{\rm rh}\gg T_w$, and (ii) a low-reheat scenario where the SM plasma is only reheated to a temperature $T_{\rm rh}\sim {\cal O}(\rm few~MeV)$ which we will fix to $T_{\rm rh}=5~\rm MeV$. This low-reheat scenario leads to relaxed bounds from the production of sterile neutrinos while also still being high enough to successfully accommodate primordial nucleosynthesis.

The production of sterile neutrino $k$ is dominantly at temperatures $T\sim{\rm min}(T_V^k,T_w)$, which we assume the SM plasma achieves in the high-reheat scenario. Equation~\ref{eq:dYdT} does not reflect the Boltzmann-suppression of the production of modes that with masses larger than this temperature. To mock up this effect, we cut off the production of modes with $M_k>{\rm min}(T_V^k,T_w)$ or, equivalently, those with
\begin{equation}
\begin{aligned}
k> k_{\rm kin}\equiv 5.1\times10^7\left(\frac{R}{10^{-4}~\mu{\rm m}}\right).
\end{aligned}
\end{equation}
Note that for all $k\lesssim k_{\rm kin}$, $T_V^k\lesssim T_w$ which means that we can ignore production for temperatures above the weak scale.

The kinematic cutoff is stricter in the low-reheat case. Modes with $M_k>T_{\rm rh}$, i.e.
\begin{equation}
\begin{aligned}
k>k_{\rm kin}^{\rm low}\simeq 2.54\times10^3\left(\frac{R}{10^{-4}~\mu{\rm m}}\right)\left(\frac{T_{\rm rh}}{5~{\rm MeV}}\right)
\end{aligned}
\label{eq:kkinlow}
\end{equation}
have exponentially suppressed production and we do not include them in our estimates. This weakens the limits drastically.

So far, we have not discussed the lifetime of the sterile states that are produced. Since we can focus on the production of states with mass below the weak scale, decays of such states are mediated by 4-Fermi weak interaction dressed by an active-sterile mixing angle,
\begin{equation}
\begin{aligned}
\tau_k&\sim\frac{\tau_\mu}{\theta_k^2}\left(\frac{m_\mu}{M_k}\right)^5\simeq 1.9\times10^{26}~{\rm s}\left(\frac{0.1~\rm eV}{m}\right)^2\left(\frac{R}{10^{-4}~\mu{\rm m}}\right)^3k^{-3}.
\end{aligned}
\end{equation}
The heaviest state that is produced has a lifetime
\begin{equation}\label{eq:heaviestlifetime}
\begin{aligned}
\tau_{k_{\rm kin}}&\sim 1.4\times10^{3}~{\rm s}\left(\frac{0.1~\rm eV}{m}\right)^2
\end{aligned}
\end{equation}
and all other states have a longer lifetime (since $k<k_{\rm kin}$). Thus, for $m\lesssim 10~{\rm eV}$, all of the states decay after SM neutrino decoupling.

We can integrate Eq.~(\ref{eq:dYdT}) in the high-reheat scenario to find
\begin{equation}
\begin{aligned}
\frac{n_k}{n_\nu}&\sim 2.32\times10^{-2} \left(\frac{m}{0.1~\rm eV}\right)^2\left(\frac{R}{10^{-4}~\mu\rm m}\right)\frac{1}{k}\left[\frac{10.75}{g_\star(T_V^k)}\right]^{11/6}.
\end{aligned}
\label{eq:nkhi}
\end{equation}
To arrive at this simple expression, we have ignored the change in the number of relativistic degrees of freedom during the relatively short temperature range where most production of mode $k$ occurs. We have also multiplied the result by a factor of $\left[10.75/g_\star(T_V^k)\right]^{4/3}$ to account for entropy injection between production and the epoch of SM neutrino decoupling.
Note the dependence of Eq. \ref{eq:nkhi} on the effective number of degrees of freedom agrees with~\cite{Luo:2020fdt}. 

In the low-reheat scenario, we will make use of the fact that for $R\lesssim 3\mu{\rm m}$, i.e. the parameter space that we are interested in, $T_V^k>T_{\rm rh}\sim 5~{\rm MeV}$. This means that during production the active-sterile mixing angle is close to its vacuum value. Integrating Eq.~(\ref{eq:dYdT}) from $T_{\rm rh}$ in this case gives
\begin{equation}
\begin{aligned}
\frac{n_k^{\rm low}}{n_\nu}&\sim 8.8\times10^{-8} \left(\frac{m}{0.1~\rm eV}\right)^2\left(\frac{R}{10^{-4}~\mu{\rm m}}\right)^2\frac{1}{k^2}\left[\frac{10.75}{g_\star(T_{\rm rh})}\right]^{11/6}\left(\frac{T_{\rm rh}}{5~{\rm MeV}}\right)^3.
\end{aligned}
\label{eq:nklo}
\end{equation}

\subsection{Contribution to the matter density}\label{sec:dm}
\begin{figure}[t]
    \includegraphics[width=0.49\linewidth]{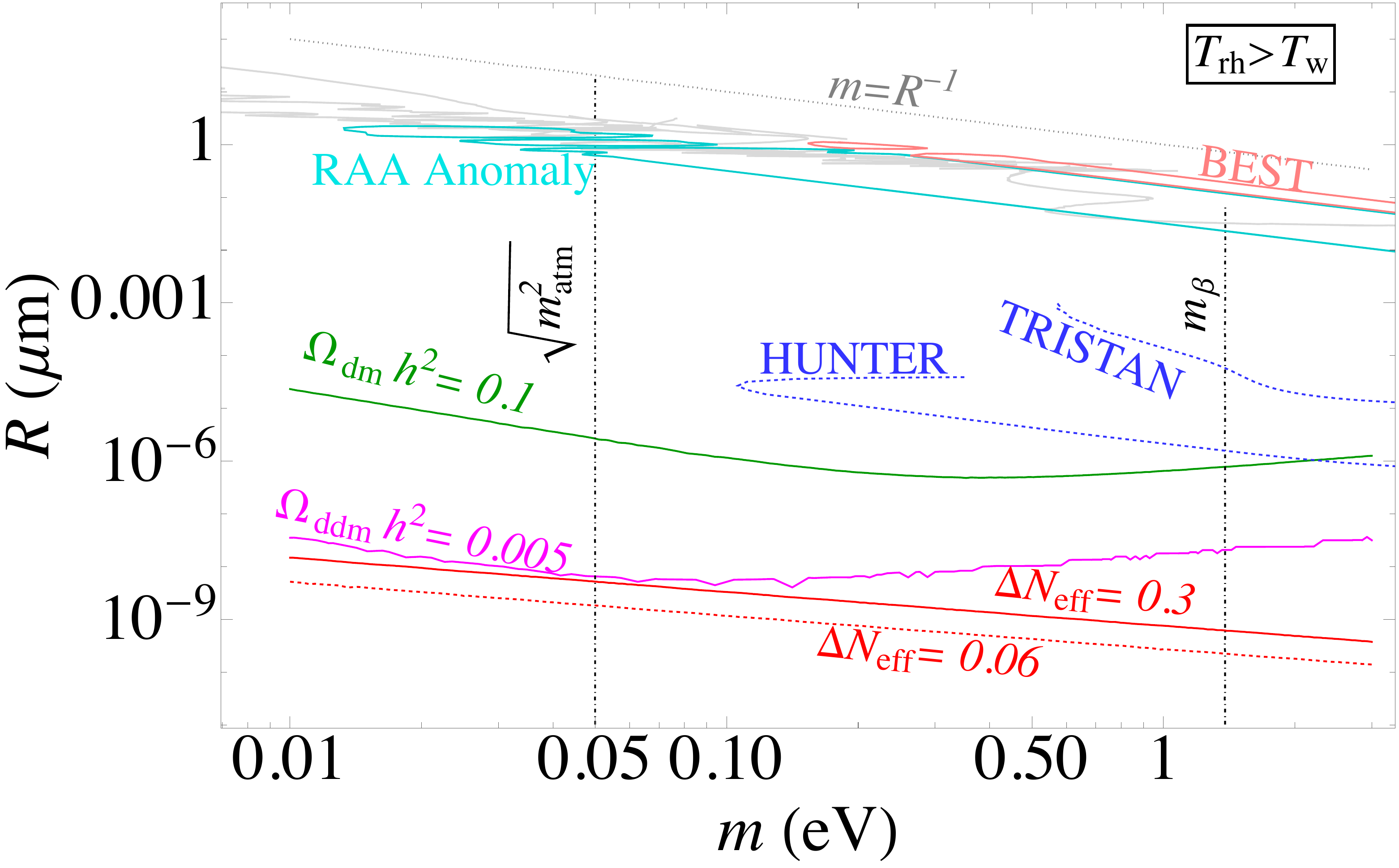}~
    \hfill
    \includegraphics[width=0.49\linewidth]{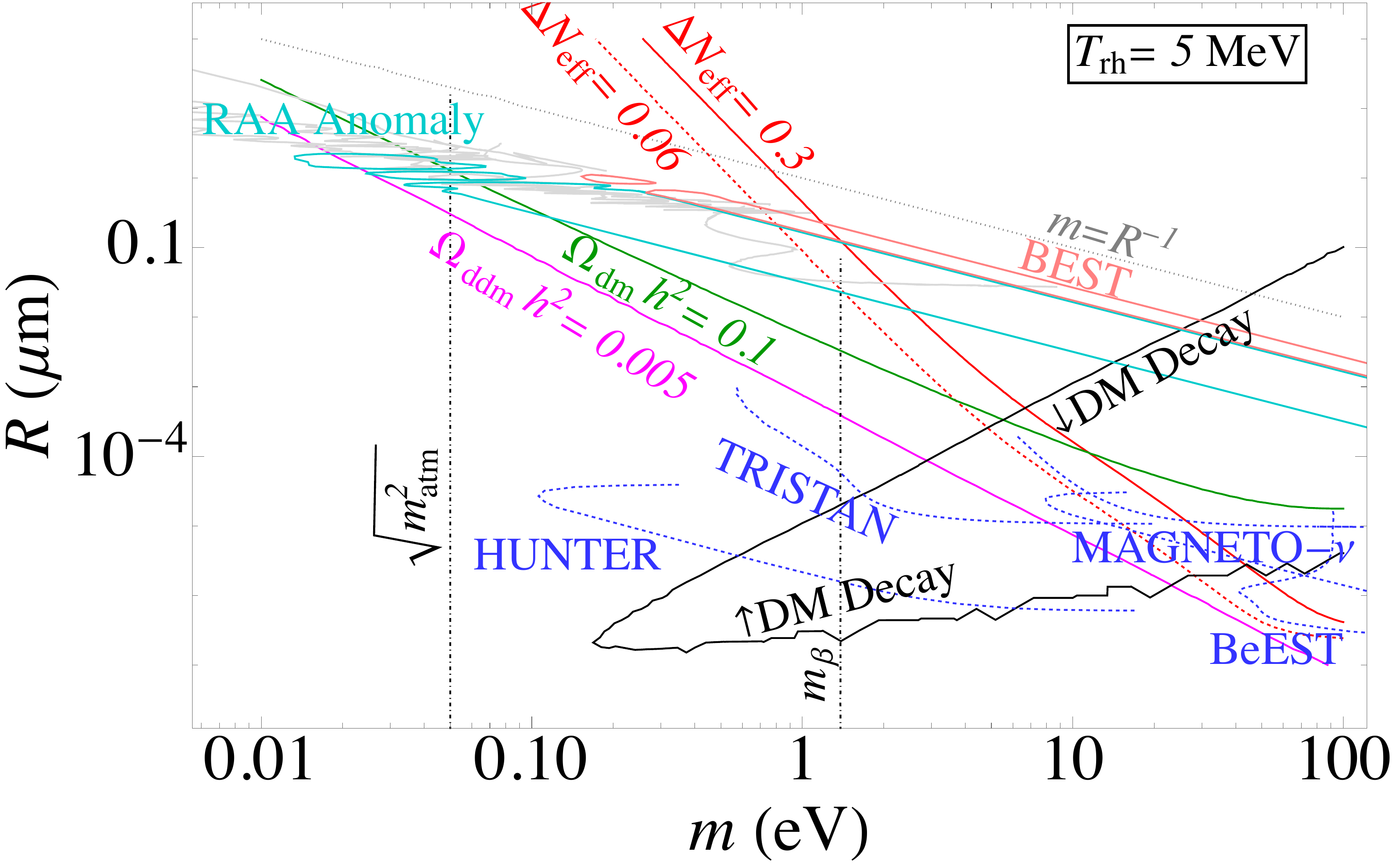}
    \caption{Upper bounds on $R$ as a function of $m$ defined in Eq.~(\ref{eq:msq}) in the high-reheat (left) and low-reheat (right) cases. We show bounds from the present limit $\Delta N_{\rm eff}<0.3$ (red, solid), not overclosing the universe (green, solid), as well as our estimate of the bound from a decaying dark matter component (magenta, solid). The region where astrophysical X-ray searches for the decay of sterile neutrino dark matter could be sensitive is within the region labeled ``DM Decay'' (black, solid). Also shown are terrestrial constraints from Fig.~\ref{fig:terrestrial} and the preferred regions for the RAA anomaly~\cite{Mention:2011rk} (teal, solid) and BEST~\cite{Barinov:2022wfh} (pink, solid). We plot projected sensitivities of the HUNTER~\cite{Smith:2016vku,Martoff:2021vxp}, MAGNETO-$\nu$~\cite{MAGNETOnu}, TRISTAN~\cite{KATRIN:2018oow}, and BeEST~\cite{Friedrich:2020nze,Leach:2021bvh} sterile neutrino search experiments (blue, dashed). See text for details.}
    \label{fig:cosmoconstraints}
\end{figure}

The presence of cold DM on cosmological scales can be distinguished from the presence of ordinary matter through measurements of the CMB angular power spectrum at small angular scales~\cite{Planck:2018vyg}. 
The best current measurements indicate DM contributes $\Omega_{\rm CDM}h^2\simeq0.1$ to the universe's overall energy density~\cite{Planck:2018vyg} where $\Omega$ represents the energy density in units of the critical energy density $\rho_{\rm cr}=10h^2~{\rm keV/cm^3}$ with $h$ related to the current Hubble expansion rate through $H=100h~{\rm km/s/kpc}$.
That said, DM need only be stable on cosmological time scales and it is possible that some fraction of what we observe as DM today has decayed since the formation of the CMB. 
If DM decays between the time of recombination and today, this will alter the growth of structure and this effect can be measured or constrained by angular anisotropies in the CMB~\cite{Poulin:2016nat,Nygaard:2020sow,Simon:2022ftd}.
The constraints on dark matter that decays to radiation are that no more than about 5\% of the dark matter density could have decayed between recombination and the present-day; this translates to $\Omega_{\rm ddm}h^2<0.005$ if the decaying component's lifetime is short compared to $t_U=13.6\times10^9~{\rm yr}$, the age of the Universe, or to the decaying dark matter's lifetime being larger than about $20t_U$ if it makes up the entirety of the dark matter energy budget.

Those KK-modes which have lifetimes $\tau_k\geq t_U$ contribute to the present-day DM density $\Omega_{\rm dm} h^2$ while those with $t_U\gtrsim \tau_k\geq t_{\rm CMB}$ contribute to a decaying DM density $\Omega_{\rm ddm}h^2$.\footnote{Sterile neutrinos are of course one of the most well studied DM candidates~\cite{Gunn:1978gr,Dodelson:1993je}. If their masses are $\sim10~\rm keV$ or smaller (corresponding to $R\gtrsim10^{-3}$) they are a warm dark matter candidate and bounds can be placed from structure formation~\cite{Bode:2000gq,Schneider:2011yu}. Determining the precise bounds when there could be multiple DM subcomponents with different masses and mixing angles is extremely complicated and beyond the scope of the present work. See, e.g.~Ref.~\cite{An:2023mkf} for a recent study on warm dark matter in a nonstandard sterile neutrino setup.} Once produced, the sterile modes have kinetic energies comparable to the SM neutrinos and free stream until the temperature is of order their mass.\footnote{The relatively small reheating of the active neutrinos by the annihilation of SM states after they become nonrelativistic does not change our results appreciably.} Since we are interested in $R\lesssim\mu{\rm m}$, even the lightest of the sterile modes is nonrelativistic at the epoch of matter-radiation equality. The energy density in these neutrinos contributes to the matter density. In the high-reheat scenario, we use Eq.~(\ref{eq:nkhi}) to write
\begin{equation}
\begin{aligned}
\rho_k\simeq M_k n_k&\sim 45.7~{\rm eV} \left(\frac{m}{0.1~\rm eV}\right)^2\left[\frac{10.75}{g_\star(T_V^k)}\right]^{11/6}n_\nu.
\end{aligned}
\label{eq:density_hi}
\end{equation}
Importantly, except for the mild dependence through the number of SM degrees of freedom at production, this expression is independent of the mass of the state $k/R$ so that each state in the KK tower that is produced contributes roughly equally. The contribution of these states to the present-day matter density in units of the critical density is
\begin{equation}
\begin{aligned}
\Omega_{\rm dm} h^2&=\sum_{k=1}^{k_{\rm kin}}\frac{\rho_k}{\rho_{\rm cr}}e^{-t_U/\tau_k}\sim \sum_{k=1}^{k_{t_U}}0.5 \left(\frac{m}{0.1~\rm eV}\right)^2\left[\frac{10.75}{g_\star(T_V^k)}\right]^{11/6}.
\end{aligned}
\label{eq:OmegaDM_HR}
\end{equation}
In the last equality above we have approximated $\exp(-t_U/\tau_k)=\theta(k_{t_U}-k)$ where
\begin{equation}
\begin{aligned}
k_{t_U}\equiv 765\left(\frac{0.1~\rm eV}{m}\right)^{2/3}\left(\frac{R}{10^{-4}~\mu{\rm m}}\right)
\end{aligned}\label{eq:ktu}
\end{equation}
is the value of $k$ such that $\tau_k=t_U$.

In addition, the sterile states can behave as a decaying dark matter component if they decay between the time of CMB formation, $t_{\rm CMB}=3.8\times10^5~{\rm yr}$, and the present day,
\begin{equation}
\begin{aligned}
\Omega_{\rm ddm} h^2&=\sum_{k=1}^{k_{\rm kin}}\frac{\rho_k}{\rho_{\rm cr}}e^{-t_{\rm CMB}/\tau_k}\left[1-e^{-t_U/\tau_k}\right]\sim \sum_{k=k_{t_U}}^{k_{t_{\rm CMB}}}0.5 \left(\frac{m}{0.1~\rm eV}\right)^2\left[\frac{10.75}{g_\star(T_V^k)}\right]^{11/6}.
\end{aligned}
\label{eq:OmegaDDM_HR}
\end{equation}
In the last step we have again made the same step function approximations of the exponentials and introduced
\begin{equation}
\begin{aligned}
k_{t_{\rm CMB}} \equiv 2.5\times10^4\left(\frac{0.1~\rm eV}{m}\right)^{2/3}\left(\frac{R}{10^{-4}~\mu{\rm m}}\right),
\end{aligned}\label{eq:ktcmb}
\end{equation}
which satisfies $\tau_{k_{\rm CMB}}=t_{\rm CMB}$.

In the low-reheat case, the energy density in mode $k$ is reduced (and $k$-dependent). From  Eq.~(\ref{eq:nklo}),
\begin{equation}
\begin{aligned}
\rho_k^{\rm low}&\sim 1.75\times10^{-4}~{\rm eV} \left(\frac{m}{0.1~\rm eV}\right)^2\left(\frac{R}{10^{-4}~\mu{\rm m}}\right)\frac{1}{k}\left[\frac{10.75}{g_\star(T_{\rm rh})}\right]^{11/6}\left(\frac{T_{\rm rh}}{5~{\rm MeV}}\right)^3 n_\nu.
\end{aligned}
\label{eq:density_low}
\end{equation}
Note that, in contrast to the high reheat case in Eq.~(\ref{eq:density_hi}), this energy density depends on the mass of mode $k$, with heavier modes contributing less. To find the contribution to the (decaying) dark matter density, we sum the contributions up to $k_{\rm kin}^{\rm low}$ in Eq.~(\ref{eq:kkinlow}) and weight each term by the appropriate factors to account for their lifetimes as done in the high-reheat case in Eqs.~(\ref{eq:OmegaDM_HR}) and~(\ref{eq:OmegaDDM_HR}).

In Fig.~\ref{fig:cosmoconstraints}, we show the upper bounds on $R$ as a function of $m$ that come from not overclosing the universe, $\Omega_{\rm dm} h^2<0.12$~\cite{Planck:2018vyg}. We also show an estimated upper bound from the decaying DM limit, $\Omega_{\rm ddm} h^2<5\%\times 0.12$~\cite{Poulin:2016nat,Nygaard:2020sow,Simon:2022ftd}. This applies in our case since the majority of the energy in the sterile neutrino decays is carried away by relativistic species. For values of $m$ larger than about $0.3\eV$ this decaying DM limit could be weakened by the slight degeneracy between the effects of nonzero light neutrino masses and decaying DM. The left panel shows the bounds in the high reheat temperature case while the right does so for $T_{\rm rh}=5~{\rm MeV}$. As we have displayed, proposed searches for sterile neutrinos, such as HUNTER~\cite{Smith:2016vku,Martoff:2021vxp}, MAGNETO-$\nu$~\cite{MAGNETOnu}, TRISTAN~\cite{KATRIN:2018oow}, and BeEST~\cite{Friedrich:2020nze,Leach:2021bvh}, can also probe some of this parameter space which we have also plotted, interpreting each as a search for the lightest KK sterile neutrino mode.

In the high reheat case, the bounds are far stronger than those from terrestrial experiments and exclude regions of parameter space that explain neutrino oscillation data anomalies. The bounds are weakened by several orders of magnitude in the low reheat case, opening up some of the neutrino anomaly parameter space; however, this region is in some conflict with the lower bound on the light neutrino masses.

The upper bound on $R$ from the observed matter density in the low reheat case, for $m\lesssim 1~{\rm eV}$, is larger than $10^{-3}~\mu{\rm m}$. In this part of parameter space, the lightest KK modes are at the $\rm keV$ scale or below, and therefore some of them act as a warm dark matter component. Although beyond the scope of the current work, studying the impact of the mixed warm and cold dark matter components on structure formation could lead to slightly tighter bounds.

Astrophysical searches for the decay products of the sterile states that survive until the present day can also be important. For sterile modes with masses of ${\cal O}(10~\rm keV)$, i.e. $R\sim10^{-4}~\mu{\rm m}$, are subject to searches using x-ray observatories such as Chandra, NuSTAR, and DEBRA~\cite{Horiuchi:2013noa,Sicilian:2020glg,Roach:2019ctw, Roach:2022lgo,Boyarsky:2005us} due to the loop-level decay to a light neutrino and a monochromatic photon. Owing to its larger number density, the $k=1$ mode dominates this signal. Generally speaking, these searches limit
\begin{equation}\label{eq:indirectconstraints}
\Omega_{k=1}h^2/\tau_{k=1}\lesssim 10^{-26}~{\rm s}^{-1}.
\end{equation}
On the right panel of Fig.~\ref{fig:cosmoconstraints}, we show the region in the low-reheat case where~(\ref{eq:indirectconstraints}) is not satisfied as a rough guide to where indirect searches for decaying dark matter are sensitive. In the high-reheat case, the region where such searches can probe is already excluded by overclosing the universe, or by the sterile neutrino states having too short a lifetime. We defer a more complete study taking into account the multiple lines from other bulk neutrino modes to future study.

\subsection{Shift of $\Delta N_{\rm eff}$}\label{sec:Neff}
Measurements of the CMB provide a very precise picture of the Universe around and after the time of matter-radiation equality, when the temperature of the SM plasma was around $0.1~\rm eV$. In particular, such measurements are very sensitive to the relativistic energy density that is not coupled to the SM plasma, such as neutrinos or other light, noninteracting species. Conventionally, this is written in terms of $N_{\rm eff}$, which is the ratio of the noninteracting relativistic energy density to that of the electromagnetically interacting plasma, $\rho_\gamma$,
\begin{equation}
N_{\rm eff}=\frac87 \left(\frac{11}{4}\right)^{4/3}\frac{\sum_\nu\rho_\nu+\rho_{\rm new}}{\rho_\gamma}=N_{\rm eff}^{\rm SM}+\Delta N_{\rm eff},~\Delta N_{\rm eff}\equiv\frac87 \left(\frac{11}{4}\right)^{4/3}\frac{\rho_{\rm new}}{\rho_\nu},
\end{equation}
where $\rho_{\rm new}$ is the contribution to the noninteracting relativistic energy density beyond the SM expectation.\footnote{Primordial nucleosynthesis is of course also sensitive to the number of relativistic degrees of freedom when the temperature of the universe was $T\sim100~\rm keV$, with the shift from the standard value typically also parameterized in terms of $\Delta N_{\rm eff}$. Modes with $M_k\lesssim1~\MeV$ contribute to this value of $\Delta N_{\rm eff}$. However, in the scenario we study, this shift is always extremely subleading when compared to the CMB value.} The prefactor is chosen so that a single thermalized neutrino species contributes $N_{\rm eff}\simeq 1$. Larger $N_{\rm eff}$ corresponds to a larger Hubble expansion rate, in a form that is free-streaming, which leads to a particular imprint on the CMB at small scales~\cite{Hou:2011ec,Bashinsky:2003tk,Follin:2015hya}. Observations from Planck presently limit $\Neff<0.3$~\cite{Planck:2018vyg}.
Upcoming cosmological surveys such as CMB Stage-IV will probe $\Neff$ to around $0.06$~\cite{CMB-S4:2016ple,Abazajian:2019eic}.

The heavy, mostly sterile neutrinos whose production we have computed above are all massive during the CMB formation epoch. Therefore they do not directly contribute to $N_{\rm eff}$. However, they decay via weak interactions through their small active admixture into standard model states including light, active neutrinos. The active neutrinos that are produced in such decays after neutrino decoupling at $T\sim{\rm MeV}$ but before the formation of the CMB contribute to $\rho_{\rm new}$, i.e. to $\Delta N_{\rm eff}$. 

To compute $\Delta N_{\rm eff}$ in this scenario, we have to track the energy density in the EM plasma as well as the new contribution in light neutrinos. If sterile mode $k$ decays after going nonrelativistic and deposits energies of $r_kM_k$ into SM neutrinos and $(1-r_k)M_k$ into the EM plasma, the evolution of the relevant energy densities is given by
\begin{align}
\frac{d\rho_{\rm new}}{dt}+4H\rho_{\rm new}&=\sum_k r_k\frac{M_k n_k}{\tau_k},
\label{eq:drhonudt}
\\
\frac{d\rho_\gamma}{dt}+4H\rho_\gamma&=\sum_k(1-r_k) \frac{M_k n_k}{\tau_k}.
\label{eq:drhogammadt}
\end{align}
The sterile neutrino number density $n_k$ is given by
\begin{equation}
\begin{aligned}
n_k=\frac{n_k(t_0)}{a^3}e^{-t/\tau_k},
\end{aligned}
\end{equation}
where $t_0$ is an early time after production but before decays become important ($t_0\ll \tau_k$) and we define the scale factor such that $a(t_0)=1$.

Integrating Eqs.~(\ref{eq:drhonudt}) and~(\ref{eq:drhogammadt}) we can write the shift in $N_{\rm eff}$ as
\begin{equation}
\begin{aligned}\label{eq:Neff}
\Delta N_{\rm eff}&\simeq \sum_k f_k \frac{M_k n_k(t_0)}{\rho_\nu(t_0)\tau_k}\int_{t_0}^{t_{\rm CMB}} dt\, a(t) e^{-t/\tau_k}
\end{aligned}
\end{equation}
with
\begin{equation}
\begin{aligned}
f_k\equiv r_k-\frac{21}{8}\left(\frac{4}{11}\right)^{4/3}(1-r_k).
\end{aligned}
\label{eq:fk}
\end{equation}
A simple accounting of available final states, including subsequent decays into neutrinos, naively gives $r_k=f_k=1$ for $M_k<2m_e$ and $r_k\simeq0.5$, $f_k\simeq0.16$ for $M_k>2m_e$.\footnote{Note that some of the electromagnetically interacting objects created in the decay of the sterile modes even before decoupling may not fully thermalize with the rest of the plasma, in this case their energy density would contribute to $\rho_{\rm new}$. The error introduced by ignoring this is only ${\cal O}(1)$ and moreover the bounds we derive are conservative. Further bounds could also come from distortions of the spectrum of CMB photons from electromagnetic energy injected after the freeze-out of (double) Compton scattering.} For the high-reheat case, in the part of parameter space where the current $\Delta N_{\rm eff}$ bound lies, all modes have $\tau_k\ll t_{\rm CMB}$ and the Universe remains radiation-dominated until the usual point of matter-radiation equality around $t_{\rm CMB}$ with $a(t)\propto\sqrt{t}$. In this case, using the freeze-in number density in Eq.~(\ref{eq:nkhi}), the shift can be approximated as
\begin{equation}
\begin{aligned}
\Delta N_{\rm eff}&\simeq \sum_{k=1}^{k_{\rm kin}} 0.103\, f_k \left(\frac{m}{0.1~\rm eV}\right) \left(\frac{R}{10^{-9}~\mu{\rm m}}\right)^{3/2}k^{-3/2}\left[\frac{106.75}{g_\star(T_V^k)}\right]^{11/6}
\\
&\simeq 0.03\left(\frac{f_k}{0.16}\right) \left(\frac{m}{0.1~\rm eV}\right) \left(\frac{R}{10^{-9}~\mu{\rm m}}\right)^{3/2}\left[\frac{106.75}{g_\star(T_V^k)}\right]^{11/6},
\end{aligned}\label{eq:neffhigh}
\end{equation}
where we have ignored the variations in $f_k$ and $g_\star(T_V^k)$ with $k$ in the last step.

In the low-reheat case, most of the calculation goes through as above but with the appropriate value of the sterile mode production from Eq.~(\ref{eq:nklo}). However, as $m$ varies, the dominant contribution to $N_{\rm eff}$ can come from modes with lifetimes smaller or larger than $t_{\rm CMB}$, which affects how the limits scale. For sterile modes with $\tau_k>t_{\rm CMB}$, the contribution of a single mode $k$ is
\begin{equation}
\begin{aligned}
\Delta N_{{\rm eff},k}^{\rm low}&\simeq 10^{-17}\,  f_k \left(\frac{m}{0.1~\rm eV}\right)^4\left(\frac{10^{-4}~\mu{\rm m}}{R}\right)^2 k^2\left(\frac{T_{\rm rh}}{5~{\rm MeV}}\right)^3\left[\frac{10.75}{g_\star(T_V^k)}\right]^{11/6},
\end{aligned}
\label{eq:deltaNlolate}
\end{equation}
while for those with $\tau_k<t_{\rm CMB}$, it is
\begin{equation}
\begin{aligned}
\Delta N_{{\rm eff},k}^{\rm low}&\sim 8.4\times10^{2}\,  f_k \left(\frac{m}{0.1~\rm eV}\right)\left(\frac{R}{10^{-4}~\mu{\rm m}}\right)^{5/2} k^{-5/2}\left(\frac{T_{\rm rh}}{5~{\rm MeV}}\right)^3\left[\frac{10.75}{g_\star(T_V^k)}\right]^{11/6}.
\end{aligned}
\label{eq:deltaNloearly}
\end{equation}

If $m\lesssim 6~{\rm eV}(5~{\rm MeV}/T_{\rm rh})^{3/2}$, all modes decay after the formation of the CMB and the sum only includes modes with $\Delta N_{{\rm eff},k}$ as in Eq.~(\ref{eq:deltaNlolate}). For $m\gtrsim 6~{\rm eV}(5~{\rm MeV}/T_{\rm rh})^{3/2}$ there are contributions from modes with lifetimes longer and shorter than the time of the CMB epoch and the total includes sums over modes with $\Delta N_{{\rm eff},k}$ of the forms in Eqs.~(\ref{eq:deltaNlolate}) and~(\ref{eq:deltaNloearly}); those with $\tau_k\sim t_{\rm CMB}$ dominate the contribution. Summarizing, we have
\begin{equation}
\begin{aligned}
\Delta N_{\rm eff}^{\rm low}&\simeq f_k\left(\frac{m}{0.1~\rm eV}\right)^2\left(\frac{R}{10^{-4}~\mu{\rm m}}\right)\left(\frac{T_{\rm rh}}{5~{\rm MeV}}\right)^3\left[\frac{10.75}{g_\star(T_V^k)}\right]^{11/6}
\\
&\times\begin{cases}
    5.3\times10^{-8} \left(\frac{m}{0.1~\rm eV}\right)^2\left(\frac{T_{\rm rh}}{5~{\rm MeV}}\right)^3, & m\lesssim6\eV \left(\frac{5~{\rm MeV}}{T_{\rm rh}}\right)^{3/2}, \\
    1.9\times10^{-4}, & m\gtrsim6\eV \left(\frac{5~{\rm MeV}}{T_{\rm rh}}\right)^{3/2}.
    \end{cases}
\end{aligned}
\end{equation}
Again, we have ignored the variation with $k$ of $f_k$ and $g_\star(T_V^k)$ to arrive at these approximate expressions.

We show the upper bounds on $R$ varying $m$ from $\Delta N_{\rm eff}<0.3$ in Fig.~\ref{fig:cosmoconstraints} for both the high reheat and $T_{\rm RH}=5~{\rm MeV}$ scenarios. Also shown is the area of parameter space that can be probed by a future measurement of $\Delta N_{\rm eff}$ as small as $0.05$ as hoped for with CMB-S4. In the high reheat scenario, this provides the strongest constraint, limiting $R\lesssim10^{-9}~\mu{\rm m}$ while in the low reheat case, it is subleading to the limits from the (decaying) dark matter density.

\section{Conclusions}\label{sec:conc}
Although extremely successful, the SM has several deficiencies, notably the lack of an explanation of the origin of neutrino masses and the omission of a quantum theory of gravity. Extra dimensions beyond the four we know of are common features of quantum gravity theories and can provide a natural understanding of the smallness of neutrino masses. Such explanations require the addition of new states that propagate in the extra dimensions and couple weakly to the active neutrinos of the standard model, leading to a KK tower of sterile neutrinos. The mass scale of the sterile neutrinos is determined by the size of the compactified extra dimensions.

From the point of view of string theory and swampland conjectures, the existence of a LED is well motivated and provides a natural explanation for the tiny masses of SM neutrinos.
Models with an extra dimension of size around a micron accommodate a tower of sterile neutrinos with masses starting at the eV scale. 
Neutrinos of this mass have been hinted at in some terrestrial neutrino experiments.
We have computed the production rate of these towers of sterile neutrinos in the early universe as functions of the size of the extra dimension. Although very weakly coupled, the towers of sterile neutrinos can be produced in large enough numbers to drastically affect cosmological observables, such as the (potentially decaying) dark matter density and the number of noninteracting degrees of freedom present at the CMB epoch. The constraints we have set correspond to the scenario where the neutrinos have Dirac masses. In the case that this model is modified to include Majorana masses, new limits and signals could of course emerge in lepton-number-violating observables, such as neutrinoless double beta decay~\cite{Bhattacharyya:2002vf}. Furthermore, such a model could potentially explain the matter asymmetry of the universe in the case of reheating high above the electroweak scale (in the low-reheating case, see our earlier discussion of models of baryogenesis that operate at low temperatures).

Currently, cosmological bounds place very strong limits on the size of the extra dimension in these models, at roughly the $10^{-9}~\mu{\rm m}$ level in the case of a standard cosmological history. In this case the leading effect is the shift in the number of noninteracting relativistic degrees of freedom, $N_{\rm eff}$. This rules out a micron-sized extra dimension as an explanation of neutrino experiment anomalies (which is also the case in most sterile neutrino explanations of such anomalies). We have explored the weakening of these limits in the case of a low reheat temperature at the MeV scale, i.e. if the SM plasma never achieved temperatures well above an MeV during its earliest stages. This is the lowest reheat temperature compatible with successful primordial nucleosynthesis. In contrast to the high reheat case, the leading bound comes from the sterile neutrinos' contributions to the matter density, particularly as a component that decays after the formation of the CMB. The bounds are substantially weakened in the low reheat case, potentially opening up an explanation of some electron neutrino disappearance anomalies. However, such an explanation could be in tension with the overall light neutrino mass scale implied by active neutrino oscillation data and limits on a decaying dark matter component. Taken together, seeing deviations from future observations that point to either a nonzero $\Delta N_{\rm eff}$ or a decaying dark matter component could help to distinguish these two scenarios from one another and to shine light on the early stages of the universe's existence.

\acknowledgments

We thank Carlos Henrique de Lima, David Morrissey, Douglas Tuckler, and Tsung-Han Yeh for helpful discussions and comments. The work of DM is supported by a Discovery Grant from the Natural Sciences and Engineering Research Council of Canada (NSERC). MS is supported by TRIUMF which receives federal funding via a contribution agreement with the National Research Council (NRC) of Canada.

\bibliography{ref}
\end{document}